\documentstyle[prd,twocolumn,aps,floats]{revtex} 
\begin{document}
\input epsf
\def\be{\begin{equation}}\def\ee{\end{equation}}
\def\bea{\begin{eqnarray}}\def\eea{\end{eqnarray}}

\def\goesas{\mathop{\sim}\limits} \def\la{\lambda} \def\etal{{\it et al.}}
\def\si{\sigma} \def\rh{\rho} \def\ph{\phi} \def\al{\alpha}
\def\bt{\beta} \def\Om{\Omega} \def\kp{\kappa}
\def\e{{\rm e}} \def\ns#1{_{\text{#1}}} \def\Omp{\Om_{\ph0}}
\wideabs{
\title{Future supernova probes of quintessence}
\author{S.C. Cindy Ng$^{\bf a}$\cite{Ecng} and
David L. Wiltshire$^{\bf a,b}$\cite{Edlw}}
\address{$^{\bf a}$Department of Physics and Mathematical Physics, Adelaide
University,\break Adelaide, S.A. 5005, Australia.}
\address{$^{\bf b}$\cite{Adlw}Department of Physics and Astronomy, University
of Canterbury,\break Private Bag 4800, Christchurch 8020, New Zealand.}

\date{7 July, 2001; ADP-01-21/M97, astro-ph/0107142; Phys. Rev. {\bf D 64},
123519 (2001).}
\maketitle

\begin{abstract}
We investigate the potential of a future supernovae data set, as might be
obtained by the proposed SNAP satellite, to discriminate between two
possible explanations for the observed dimming of the high redshift type
IA supernovae, namely either (i) a cosmological evolution for which the
expansion of the universe has been accelerating for a substantial range of
redshifts $z\goesas1$; or (ii) an unexpected supernova luminosity evolution
over such a redshift range. By evaluating Bayes factors we show that
within the context of spatially flat model universes with a dark energy
the future SNAP data set should be able to discriminate these two
possibilities. Our calculations assume particular cosmological models with
a quintessence field in the form of a dynamical pseudo Nambu--Goldstone boson
(PNGB), and a simple empirical model of the evolution of peak luminosities
of the supernovae sources which has been recently discussed in the literature.
We also show that the fiducial SNAP data set, simulated with the assumption
of no source evolution, is able to discriminate the PNGB model from a number
of other spatially flat quintessence models which have been widely studied
in the literature, namely those with inverse power--law, simple exponential
and double--exponential potentials.
\end{abstract}
}
\narrowtext

\section{Introduction}

Type Ia supernovae (SNe Ia) can be used as standard candles to infer the
luminosity distance, $d_L$, as a function of redshift
\cite{Per97,Rie98,Per99}, and
such data provide a key element in the case for cosmic acceleration.
The measurements provide a simple way to estimate cosmological parameters.
This is the aim of at least two collaborations: the Supernova Cosmology
Project \cite{Per97,Per99} and the High-redshift Supernova Search Team
\cite{Rie98}. However, the analysis of some $\sim50$ supernovae at
redshifts around $z\goesas0.5$ has not yet provided strong constraints
on the nature of the ``dark energy''.

``Quintessence'' \cite{Wet,PR,CDS} has been proposed as a candidate for the
``dark energy'' to provide a dynamical solution to the cosmological
constant problem \cite{Bin}. The supernova data, although sufficient to
constrain the parameters of quintessence models\cite{WM,Ng,NW}, is not yet
abundant enough to allow a discrimination between competing quintessence
models. Moreover, the unexpected dimming of the type IA supernovae at redshifts
$z\goesas0.5$ is a phenomenon which could be readily attributed either to an
accelerating Universe or to an unexpected luminosity evolution \cite{Rie99}.
The present data alone is not sufficient to allow a discrimination between
these two possibilities with complete confidence \cite{NW,Wi,DLW}.

Very recently the identification of a SN Ia event at redshift $z\goesas1.7$
\cite{Rie01} has provided tantalizing evidence that the expansion of the
universe could have been decelerating at that epoch. Naturally, no firm
conclusions can be drawn from a single event. However, if this result holds
up it would rule out the simplest models of source evolution. Furthermore,
supernovae events at such redshifts are the type of observations which
should enable a discrimination to be made between some of the different
quintessence models.

Much more data is needed to enable an accurate estimation of the nature of
the dark energy. This might be accomplished by a dedicated space telescope,
the SuperNova Acceleration Probe (SNAP) \cite{snap,pencil}, which aims to
collect a large number of supernovae with $z<2$. In this paper, we assess the
ability of the SNAP mission to determine various properties of the ``dark
energy''. By analysing a simulated data set, as might be obtained by the
proposed SNAP satellite, we can test the ability of such experiments to
distinguish among currently attractive quintessence models
\cite{Wet,PR,Ng,NW,HR,FHSW,exp2,power2,dsb,St,SW}.

The feasibility of determining the properties of the dark energy component by
using simulated data sets has already been considered by several authors
\cite{St,Hut,Maor,WA,BM,WG}. One common approach is to assume that the
quintessence field is described by a perfect fluid with equation of state
$P=w\rh$, where $w$ is approximately constant over epochs of interest, or else
slowly varying. For example, various authors \cite{Maor,WA,BM,WG} consider
models with an equation of state linear in redshift, $w(z)=w_0+w_1z$.
However, many realistic cosmological models could fall outside the confines
of these approximations. In this paper we wish to take a different
approach by considering models in which the quintessence field is
directly given by a Lagrangian, and in which $w(z)$ has the freedom to
vary widely over measurable redshifts. We will fit simulated SNAP data sets
to the exact $d_L(z)$ of different quintessence models derived from
numerical integration of the coupled Einstein-scalar field equations.

For the fiducial cosmological model used in the simulation of the supernova
data set, we consider a form of quintessence, an ultra-light pseudo
Nambu--Goldstone boson (PNGB) \cite{HR} which is still relaxing to its
vacuum state. Our reasons for this choice of quintessence model are twofold.
Firstly, from the viewpoint of quantum field theory PNGB models are the
simplest way to have naturally ultra-low mass, spin-$0$ particles and hence
perhaps the most natural candidate for a presently-existing minimally-coupled
scalar field. The effective potential of a PNGB field $\ph$ can be taken to
be of the form \cite{FHSW}
\be\label{VPNGB}V(\ph)=M^4[\cos(\ph/f)+1]\ ,\ee
where the constant term ensures that the vacuum energy vanishes at the
minimum of the potential. This potential is characterized by two mass scales,
a purely spontaneous symmetry breaking scale $f$ and an explicit symmetry
breaking scale $M$.

The second motivation for choosing a PNGB model rather than other forms
of quintessence is that it provides a natural framework for studying the
question of the possibility of source evolution.

Analyses of the SN Ia data performed with source evolution have been undertaken
in the case of open Friedmann--Robertson--Walker (FRW) models \cite{DLW}.
However, the measurement of the positions of the first acoustic peaks in the
angular power spectrum of cosmic microwave background radiation anisotropies
by the BOOMERANG, MAXIMA and DASI experiments now gives unequivocal evidence
that the Universe is close to being spatially flat \cite{boom,max,dasi}. This
makes the choice of a spatially open FRW model uncompelling.

On the other hand, the PNGB cosmologies have the virtue that while they are
spatially flat, there is no a priori preference for either accelerated
or decelerated expansion \cite{NW,Wi}.
Both possibilities are available at modest
redshifts, depending on parameter values. Ultimately, the scalar field will
undergo coherent oscillations about its minimum, and the resulting luminosity
distance will become indistinguishable from that of an Einstein--de Sitter
model at late times, although
the fraction of energy density in clumped matter, $\Omega_{m}$, and the
fraction of energy density in quintessence, $\Omega_{\ph}$, can take
any values consistent with $\Omega_{m}+\Omega_{\ph}=1$. Whether the scalar
field is currently rolling down the potential for the first time in the
history of the universe -- leading to a luminosity distance relation similar
to that produced by a cosmological constant -- or whether it has already
undergone one or more oscillations by the present epoch, is a matter of
a choice of initial conditions. Beyond some bounds set by primordial
nucleosynthesis these initial conditions are not very much constrained by
our present knowledge of the models, resulting in diverse possibilities for
cosmological evolution. The current and final values of
$\Omega_{m}$ and $\Omega_{\ph}$ likewise depend on initial conditions for
the scalar field, and on the parameters $M$ and $f$.

\section{accelerating Universe vs luminosity evolution}

In an earlier paper \cite{NW}, we considered the observational constraints
arising from SNe Ia data and gravitational lensing data
on cosmological models based on Einstein gravity minimally coupled to a
scalar quintessence field with a PNGB potential (\ref{VPNGB}). In the
case of the supernovae, we studied the constraints on the PNGB parameters
$M$ and $f$ from 60 supernovae data from the Supernova Cosmology Project
(hereafter P98) of Perlmutter \etal\ \cite{Per99}. We numerically evolved the
coupled Einstein-scalar field equations of motion forward from the epoch
of matter-radiation equality to obtain the luminosity distance -- redshift
relation, comparing the results with the data both with and without an
allowance for the possibility of source evolution.

In allowing for source evolution we are acknowledging that the peak
luminosities of distant SNe Ia have been normalized according to empirical
``Phillips relations'' \cite{Ph93}--\cite{R96} between
observed peak luminosity and supernova decay time, which have been found to
be valid at low redshifts. Although we would hope that such relations
remain applicable at high redshifts, until the Phillips relations can be
modelled and accounted for physically some doubt will always remain about
applying these relations at higher redshifts.

A simple empirical model for possible source evolution was employed in ref.\
\cite{NW}: following Drell, Loredo and Wassermann \cite{DLW} we added a term
$\bt\ln(1+z)$ to the distance modulus. This particular luminosity evolution
function is simply chosen as an illustrative example, and is not singled out
by any particular physical model. Naturally, one can criticize it on these
grounds. However, the Phillips relations are also purely empirical, and the
purpose of our study as with that of ref.\ \cite{DLW} was simply to test the
extent to which any form of source evolution was able to account for the
observed data as compared to an accelerated expansion.

\begin{figure*}[htp] \centering \leavevmode\epsfysize=9cm
\epsfbox{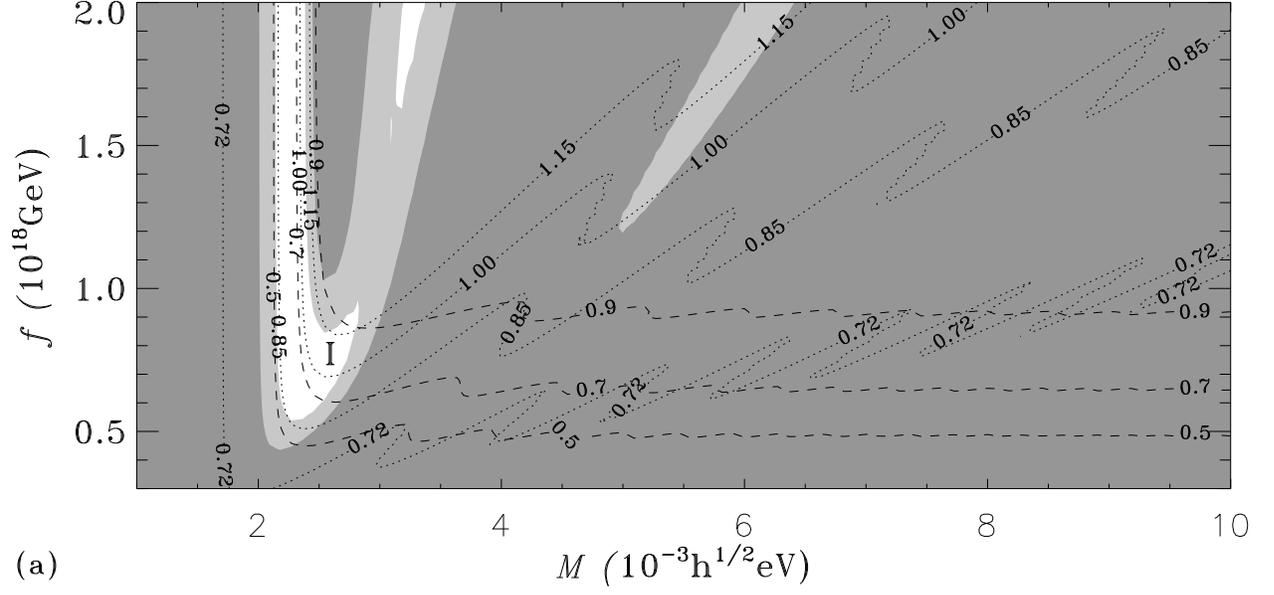} \vskip-10pt
\epsfysize=9cm \epsfbox{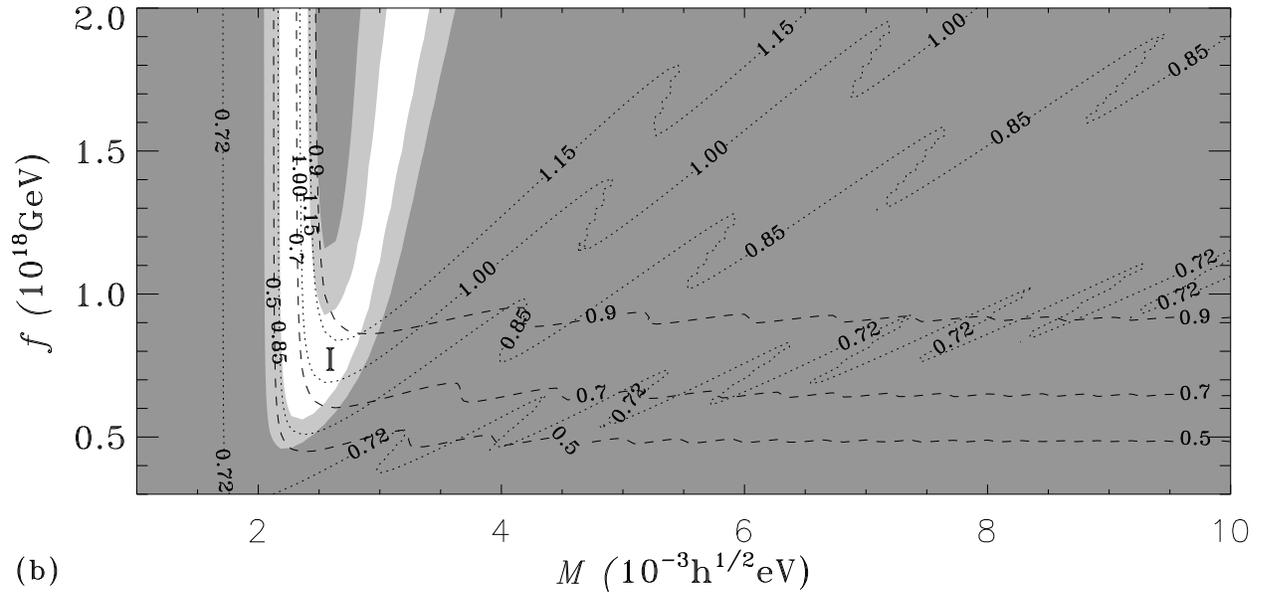}
\caption[sne60]{\label{sne60}
Confidence limits on $M$,$f$ parameter values, with $w_i=0.2$:
{\bf(a)} for the 60 supernovae Ia in the P98 data set; {\bf(b)} for the
54 supernovae Ia in the reduced ``fit C'' P98 data set.
Parameter values excluded at the 95.4\%
level are darkly shaded, while those excluded at the 68.3\% level are
lightly shaded. Overplotted are the contours for $\Omp$ (dashed) and
$H_0t_0$ (dotted). } \end{figure*}

For the purposes of analysing the data it was assumed in refs.\
\cite{NW,Wi,DLW} that the prior for the parameter $\bt$ was a Gaussian with
a mean $\bt_0$ and standard deviation $b$. The best--fit value of $\bt_0$
for the P98 data set was found to vary very slightly with initial
conditions, taking a value $\bt_0=0.414$ for $w_i\equiv\ph_i/f=1.5$ and
$\bt_0=0.435$ for $w_i=0.2$, where $\ph_i$ is the value of the quintessence
field at the time of onset of matter domination ($z=1100$). In this paper
we use the initial value $w_i=0.2$. Our reason for doing this is that in
general, and especially for larger values of $w_i$, there is a tension
between the values of $\Omp$ and $H_0t_0$ favoured by current
observations, with larger values of $H_0t_0$ often corresponding to
values of $\Omp$ rather greater than 2/3. It is still possible to find
parameter values of $M$ and $f$ which give $\Omp\goesas0.7$ while
yielding an age for the Universe consistent with the lower bound of the
current observational range \cite{La}, even for $w_i=1.5$. However, one
finds that the tension between the values of $\Omp$ and $H_0t_0$ is more
easily mitigated by choosing a lower value of $w_i$, for which the
scalar field spends more of its early dynamical history higher up the
potential hill. Confidence limits on the $M,f$ parameter space obtained
from the P98 data set, assuming no evolution, are shown in Fig.~\ref{sne60}.

In refs.\ \cite{NW,Wi} we included all 60 data from the P98 data set. In
fact, in ref.\ \cite{Per99} Perlmutter \etal\ considered a subset of 54
SNe Ia which excluded 6 supernovae events: two that are the most
significant outliers from the average light--curve width, two that have
the largest differences between observed and expected magnitudes (or
fluxes), and two that are likely to be reddened. It was discovered that
excluding these six supernovae produced a more robust fit of the cosmological
parameters $\Om_{m0}$ and $\Om_{\Lambda0}$. In refs.\ \cite{NW,Wi} we took
a conservative approach and included all 60 SNe Ia, since we are investigating
models with different cosmological parameters from those fitted in ref.\
\cite{Per99} and we did not wish to prejudge matters. However, one may of
course redo the analysis of refs.\ \cite{NW,Wi} for the ``fit C'' data set of
Perlmutter {\it et al}. For the PNGB model we find that in the best--fit
case this reduces the normalized $\chi^2$ parameter from 101.6, i.e., 1.75
per degree of freedom, for the full data set to 58.65, or 1.13 per
degree of freedom, for the reduced ``fit C'' data set. For the models
with source evolution the normalized $\chi^2$ is similarly reduced from
101.1, or 1.74 per degree of freedom, to 58.21, or 1.12 per degree of freedom.
Thus the ``fit C'' data provides a more robust fit in all cases we have
studied, whatever the model assumptions.

Since the results in ref.\ \cite{NW,Wi} were based on the full P98 data
set, we have included results based on both the full data set and the
reduced data set in Fig.~\ref{sne60}, to compare the differences. Note
that although the 2$\si$ allowed area to the right disappears entirely
for the reduced data set for $w=0.2$, this is not the case for $w=1.5$,
and thus the discrepancy between the P98 data set and that of Riess \etal\
\cite{Rie98} which was noted in Ref.\ \cite{NW} remains for the PNGB
model whether or not one uses the full or reduced data set.

\begin{figure*}[htp] \centering \leavevmode\epsfysize=9cm
\epsfbox{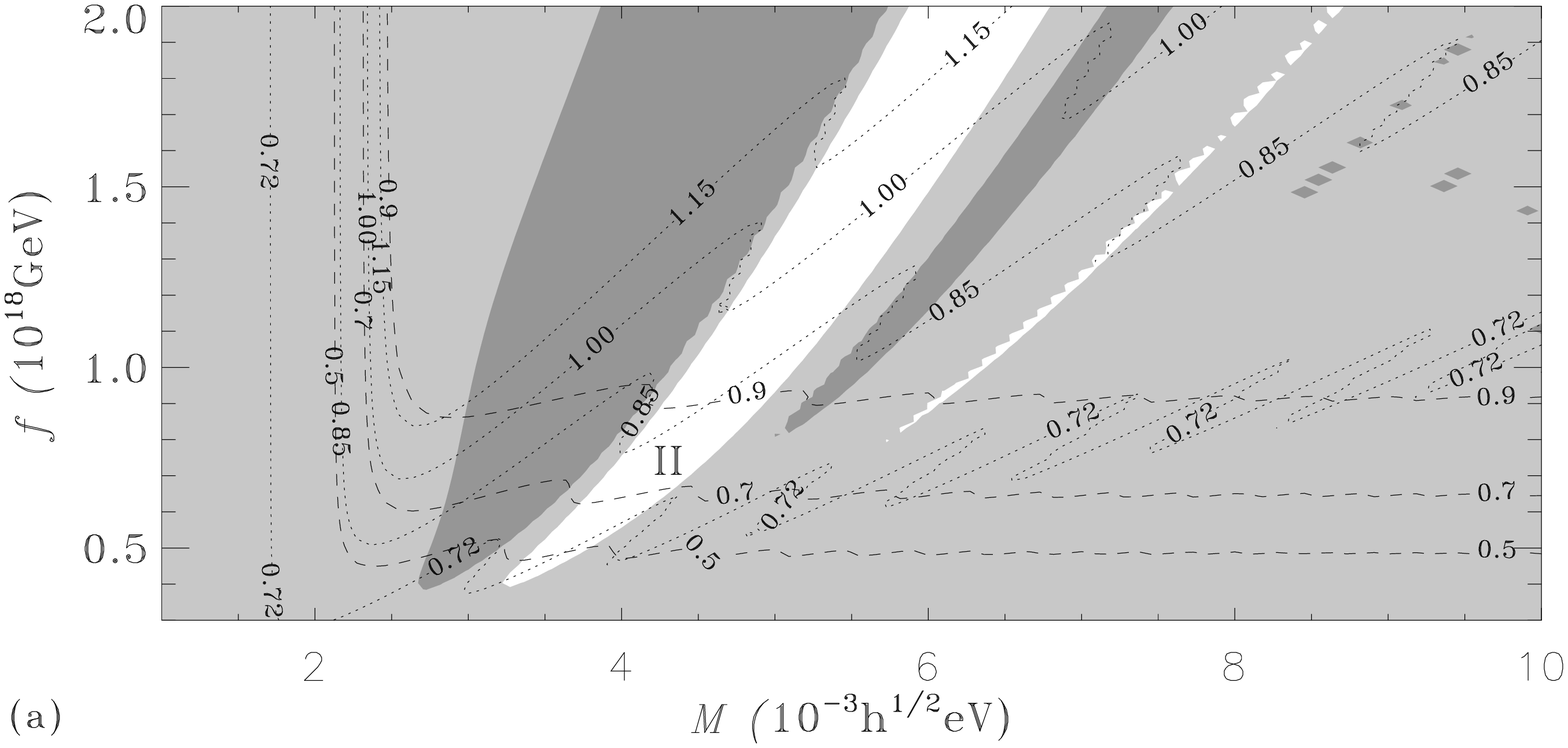} \vskip-10pt
\epsfysize=9cm \epsfbox{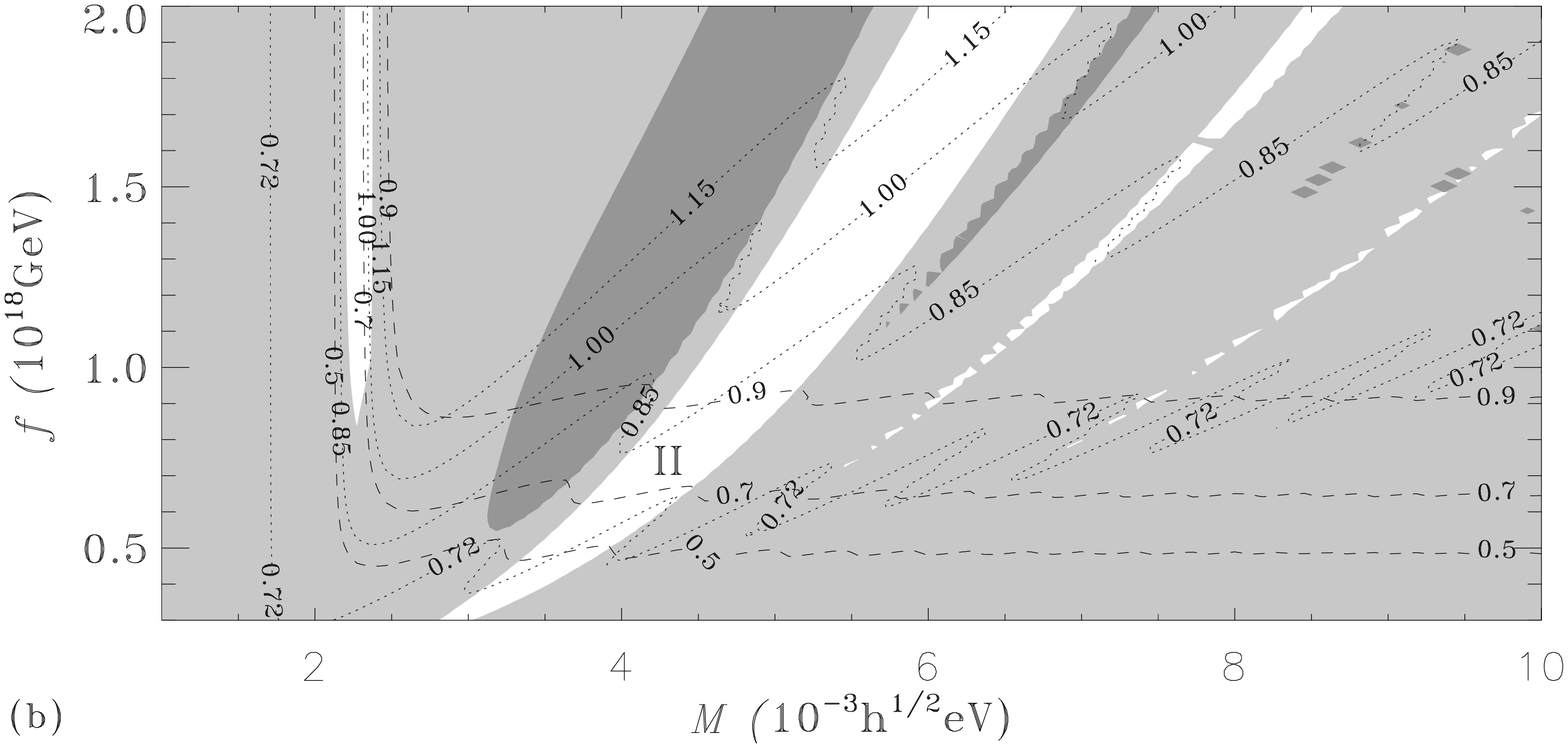}
\caption[sne60-evoln]{\label{sne60-evoln} Confidence
limits on $M$,$f$ parameter values, with $w_i=0.2$, marginalized over a
flat prior for the luminosity evolution parameter $\bt$: {\bf(a)} for
the 60 supernovae Ia in the P98 data set; {\bf(b)} for the 54 supernovae
Ia in the reduced ``fit C'' P98 data set. Parameter values excluded at
the 95.4\% level are darkly shaded, while those excluded at the 68.3\%
level are lightly shaded. Overplotted are the contours for $\Omp$ (dashed) and
$H_0t_0$ (dotted).} \end{figure*}

Here we will take a somewhat different approach to ref.\ \cite{NW} in our
treatment of the case with source evolution. We will assume that the parameter
$\bt$ has a flat prior bounded over some range $\Delta\bt=2.3$, with
limits corresponding to $-0.6<\bt<1.7$, and marginalize the likelihood
function over this prior. The choice of the bounds on $\bt$ in
explained below in section IIB.

Despite the differences from the approach used in Ref.\ \cite{NW}
the conclusions of the analysis are broadly similar, and are shown in
Fig.~\ref{sne60-evoln}. In particular, two regions of parameter space appear
to be singled out: region I (see Fig.~\ref{sne60}), which corresponds to
cosmological models for which the scalar field is rolling down the potential
(\ref{VPNGB}) for the first time (with $\ph$ increasing) at the present epoch,
and region II (see Fig.~\ref{sne60-evoln}) in which the scalar field is rolling
down the potential for the second time (with $\ph$ decreasing) at the present
epoch. In the absence of any evolution the P98 data set favours Region I,
whereas if evolution is allowed for then region II appears to be favoured
slightly more than region I.

Note that over the entire parameter space considered, we find the value
of $\hat\bt$ with the best fit of all is $\hat\bt\approx0.622$ using the
entire P98 data set, which occurs at parameter values $M=4.55\times10^{-3}
h^{1/2}$eV, $f=0.884\times10^{18}$GeV, while for the ``fit C'' data set it is
$\hat\bt\approx0.720$, which occurs at $M=4.64\times10^{-3}h^{1/2}$eV,
$f=0.901\times10^{18}$GeV. Averaged over the $(M,f)$ parameter space we find
$\langle\hat\bt\rangle\approx0.286$ for the full P98 data set, and $\langle\hat
\bt\rangle\approx0.348$ for the ``fit C'' data set. Thus in both cases the
value of $\hat\bt$ is peaked in region II, but for the ``fit C'' data
the overall values of $\hat\bt$ are somewhat larger.

The results of refs.\ \cite{NW,Wi} were inconclusive. Ostensibly the source
evolution models appeared to provide a slightly better fit. However, the
extent to which they were better was not quantified statistically. One
relatively straightforward way to compare the relative strengths of
rival theoretical models in fitting a common set of data is by the
calculation of Bayes factors \cite{DLW,KR}, even when the models involve
different numbers of parameters. Given more parameters, one may always
find a better fit to the data -- however, the Bayes factor approach
includes an effective ``Ockham's razor factor'' which adds a weighting
against the inclusion of extra parameters. The Bayes factor method may be
readily applied to the results of \cite{NW,Wi} to give an explicit statistic
to compare the relative strengths of a cosmological explanation of the
SNe Ia data versus an explanation in terms of source evolution.

In Bayesian inference, when comparing rival models $M_i$, each with
parameters ${\cal Q}_i$, the likelihood for a model conditional on the data
($D$) in a model comparison
calculation is equal to the average likelihood for its parameters,
\be
p(D|M_i)=\int d{\cal Q}_i\mbox{ }p({\cal Q}_i|M_i){\cal L}({\cal Q}_i)\ ,
\ee
where $p({\cal Q}_i|M_i)$ is the prior probability for ${\cal Q}_i$, and
${\cal L}({\cal Q}_i)$ is the sampling probability for $D$ presuming $M_i$
to be true. The ratio of model likelihoods,
\be
B_{ij}={p(D|M_i)\over p(D|M_j)}
\ee
is called the {\it Bayes factor}. When the prior odds
does not strongly favour one model over another, the Bayes factor can be
interpreted just as one would interpret an odds in betting.
Kass and Raftery \cite{KR} provide a comprehensive review of Bayes factors,
and the recommended interpretation is summarized in Table 1.
\begin{table}[hb]
\centering
\caption{Interpretation of Bayes Factors}
\bigskip
\begin{tabular}{l|l}
$B_{ij}$ & Strength of evidence for $H_i$ over $H_j$ \\
\hline
1 to 3 & Not worth more than a bare mention\\
3 to 20 & Positive\\
20 to 150 & Strong\\
$>150$ & Very Strong\\
\end{tabular}
\end{table}

\subsection{P98 data set}

In order to ascertain what degree of improvement could be expected with
data from SNAP, we will begin by calculating Bayes factors for the
existing P98 data set.

The prior ranges for parameters play an important role in Bayesian model
comparison. We will make a choice of the prior ranges of parameters in the
different models by the resulting values of $\Omega_{m0}$ and $H_0t_0$ that
they give rise to. In particular, we will choose a conservative bound of
$0.5\le\Omp\le0.9$. For the expansion age of the universe we
choose $0.72\le H_0t_0\le1.15$: for $H_0=70\mbox{ km s}^{-1}\mbox{Mpc}^{-1}$
it corresponds to a conservative bound of $t_0\simeq13\pm3$ Gyr consistent
with recent estimates \cite{La}. The parameter space bounded by these two
constraints is indicated by the contours of $\Omega_{m0}$ and $H_0t_0$ in
Figs.~\ref{sne60} and \ref{sne60-evoln}. In the large $f$ region in region I
where the contours appear to be parallel to the $f$--axis, a cut-off at
$f\goesas10^{19}$ GeV, the Planck scale, is chosen.

We calculate the average parameter likelihood by integrating ${\cal L}(M,f)$
over the prior ranges for the parameter and dividing the integral by the
relevant area. For the full P98 data set the Bayes factor for
the model without luminosity evolution versus the model with luminosity
evolution, which is the ratio of the average parameter likelihoods, is
$B\simeq2.6$. For the reduced ``fit C'' data set we obtain
a Bayes factor of 2.5, which is the same as far as its interpretation is
concerned. Thus the data alone cannot discriminate between the two
hypotheses.

\subsection{SNAP data sets}

We will now simulate a data set which would be expected to be
obtained by the SNAP satellite to investigate its potential for discriminating
between the two possibilities. We will simulate two data sets from fiducial
cosmologies chosen with parameters in each of regions I and II. In each case
we will introduce a random error to the exact distance moduli to
simulate a future supernova data set that
has been converted to a table of effective magnitudes $m_i$ and redshifts
$z_i$ of objects with a single fiducial absolute magnitude ${\cal M}_0$.
We will consider both statistical and systematic
uncertainties in the magnitudes. Typically the redshifts are known to
sufficiently high precision that their uncertainties can be ignored.
We assume the supernovae observed uniformly within four different redshift
ranges with the following different sampling rates, which are the same as
those chosen by Weller and Albrecht \cite{WA}: In the first range from
$z=0$--$0.2$ we assume that there are $50$ observations, in the second and
largest redshift range from $z=0.2$--$1.2$ there are $1800$ SNe and in the
two high redshift bins, $z=1.2$--$1.4$ and $z=1.4$--$1.7$, there are $50$
SNe and $15$ SNe observations respectively. The statistical error in
magnitude is assumed to be $\si\ns{mag} = 0.15$, including both
measurement error and any residual intrinsic dispersion after calibration.

For data set A, we assume there is no luminosity evolution. We choose the
fiducial parameters $M=2.53\times10^{-3}h^{1/2}$eV and $f=0.58\times10^
{18}$GeV; these parameters give $(\Omega_{m0},\Omp)\approx(0.23,0.77)$ and
$H_0t_0\approx0.9$. For data set B, we assume there is a luminosity
evolution and we add a term $\bt\ln(1+z)$ to the distance moduli.
We choose the fiducial parameters $M=4\times10^{-3}h^{1/2}$eV
and $f=0.676\times10^{18}$GeV. These parameters give
$(\Omega_{m0},\Omp)\approx(0.25,0.75)$ and $H_0t_0\approx0.8$. The
best--fit value of the parameter $\bt$ for this particular choice of $M$
and $f$ is $\hat\bt=0.659$ for the full data set, and $\hat\bt=0.727$
for the ``fit C'' reduced data set. We will take $\bt\approx0.659$ as the
fiducial parameter for data set B, as it is the somewhat more
conservative value, representing less luminosity evolution.

The fiducial parameters for the two data sets have been deliberately chosen
so as to give similar values of $\Omega_{m0},\Omp$ and $H_0t_0$ on one
hand, so as to be consistent with values favoured by other observational
tests. On the other hand, the fiducial parameters for data set A are
centred in region I, which corresponds to a universe in which the scalar
field is rolling down the potential for the first time at the present
epoch, whereas the fiducial parameters for data set B are centred in
region II, which corresponds to a universe in which the scalar field is
rolling down the potential for the second time at the present epoch.
As was discussed in \cite{NW} region I is favoured if the SNe Ia
luminosity distances are true cosmological distances, whereas region II
becomes significantly favoured if the data hides a simple luminosity
evolution.

We will use Bayesian analysis to estimate uncertainties. Rigorous calculation
of the likelihood for the quintessential parameters ${\cal Q}$ is very
complicated, requiring the introduction and estimation of many additional
parameters, including parameters from the light--curve model and parameters
for characteristics of the individual supernovae. With several simplifying
assumptions \cite{DLW}, the final result is relatively simple. One finds
\be
{\cal L}({\cal Q})\simeq\e^{-\chi^2/2}\ ,
\ee
where
\be
\label{chisq}
\chi^2({\cal Q})=\sum_i{[\mu_{s,i}-\mu(z_i;{\cal Q})]^2\over
\si_{\mu,i}^2}\ .
\ee
In the above equation,
\be
\mu(z_i;{\cal Q})=5\log d_L(z_i;{\cal Q})+25 \
\ee
is the distance modulus predicted by each model with parameters ${\cal Q}$,
while $\mu_{s,i}=m_i-{\cal M}_0$ is the simulated distance modulus, and its
uncertainty is $\si_{\mu,i}=0.15$. Note that we fix Hubble parameter to
$H_0=70 \mbox{ km/s Mpc}^{-1}$ to simplify the computation.

The model with an unexpected luminosity evolution corresponds to replacing
eq.~(\ref{chisq}) with
\be
\chi^2({\cal Q},\bt)=\sum_i{[\mu_{s,i}-\bt\ln(1+z_i)-\mu(z_i;{\cal Q})]^2
\over\si_{\mu,i}^2}\ .
\ee
We will marginalize over $\bt$ with a flat prior, to obtain the marginal
likelihood
\be
{\cal L}({\cal Q})={1\over\Delta\bt}\int d\bt\e^{-\chi^2/2}\ .
\ee

The above integration can be performed by an analytic marginalization
\cite{NW,DLW}. We separate $q$ from $\chi^2$ where
\bea
q({\cal Q})&=&-{\hat\bt^2({\cal Q})\over\bar\si^2}
+\sum_i{h^2(z_i;{\cal Q})\over\si_i^2} \nonumber \\
&=&\sum_i{[h(z_i;{\cal Q})-\hat\bt({\cal Q})\ln(1+z_i )]^2\over
\si_{\mu,i}^2}
\eea
is independent of $\bt$. The integral is thus an integral over a Gaussian
in $\bt$ located at $\hat\bt$ with standard deviation $\bar{\si}$.
$\hat\bt$ is the best--fit value of $\bt$ given ${\cal Q}$, and
$\bar{\si}$ is its conditional uncertainty. They are given by
\bea
{1\over\bar\si^2}&=&\sum_i{[\ln(1+z_i)]^2\over\si_{\mu,i}^2}\ , \\
\hat\bt({\cal Q})&=&\bar\si^2\left[\sum_i{h(z_i;{\cal Q})\ln(1+z_i)
\over\si_{\mu,i}^2}\right]\ ,
\eea
where
\be
h(z_i;{\cal Q})=\mu_{s,i}-\mu(z_i;{\cal Q})\ .
\ee
As long as $\bt$ is inside the prior range and $\bar{\si}\ll\Delta\bt$,
the value of the integral is well approximated by $\bar\si\sqrt{2\pi}$, so
that
\be
{\cal L}({\cal Q})={\bar\si\sqrt{2\pi}\over\Delta\bt}\e^{-q/2}\ .
\ee

The question of what bounds to place on the prior range for the parameter
$\bt$ is an interesting one, since we are dealing with a purely empirical
model of source evolution with no a priori restrictions on $\bt$, other than
that very large values of $\bt$ would not be physically plausible. We will
therefore seek the narrowest range of values of $\bt$ consistent with our
numerical results. One condition on the prior range for $\bt$ is
that $\hat\bt$ also lies inside the prior range: setting too tight a
bound on $\Delta\bt$ would count against parameter values of $M$ and $f$
which might otherwise be included. By explicit
numerical integration we have found that for the initial conditions
and the range of parameter values of $M$ and $f$ we have considered $\hat\bt$
lies in the range $-0.6<\hat\bt<1.7$. We will therefore take these
bounds to be the prior range for $\bt$.

\begin{figure*}[htp] \centering \leavevmode\epsfysize=9cm
\epsfbox{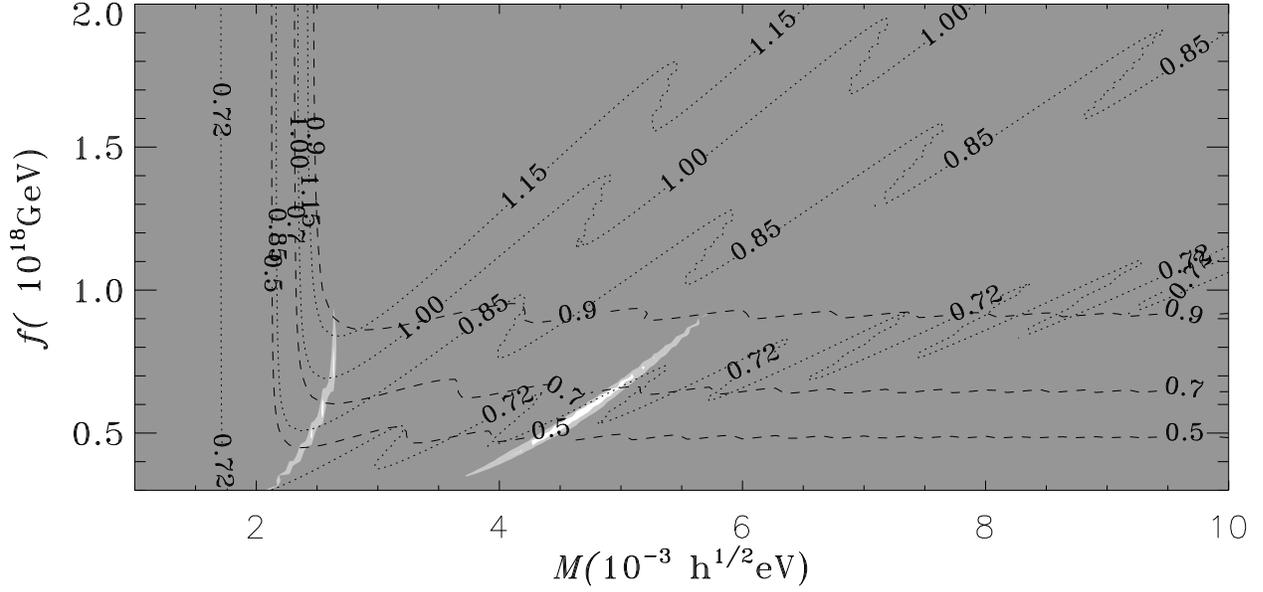} \caption[snap]{\label{snap} Confidence
limits on $M$,$f$ parameter values, with $w_i=0.2$, marginalized over a
flat prior for the luminosity evolution parameter $\bt$, for the 1915
supernovae simulated assuming $M=2.53\times10^{-3}h^{1/2}$eV,
$f=0.58\times10^{18}$GeV, and $\bt=0$.
Parameter values excluded at the 95.4\%
level are darkly shaded, while those excluded at the 68.3\% level are lightly
shaded. Overplotted are the contours for $\Omp$ (dashed) and
$H_0t_0$ (dotted).} \end{figure*}

If we fit the data set obtained from fiducial model A assuming that there is
no luminosity evolution, the parameters $M$ and $f$ are well constrained by
the simulated data. The 95.4\% confidence level contour bounds a very small
region around the best--fit values of $M\approx2.55\times10^{-3}h^{1/2}$eV and
$f\approx0.592\times10^{18}$GeV, with $\chi^2=1910$. These best--fit
parameters are very close to the fiducial parameters that
generate the data set. The average parameter
likelihood is ${\cal L}\ns{ave}\simeq7.1\times10^{-4}{\cal L}_0$, where
${\cal L}_0=\exp(-1910/2)$. If luminosity evolution is considered to exist
(see Fig. \ref{snap}), fitting the data set from fiducial model A we find
that the 95.4\% confidence level contour includes a larger region in region I
of the parameter space around the fiducial parameters, and a region in region
II. The average parameter likelihood is ${\cal L}\ns{ave}\simeq1.9\times10^
{-3}{\cal L}_0$. Hence, the Bayes factor for the model with luminosity
evolution versus the model without luminosity evolution is $B\simeq2.7$. We
conclude that it is not possible to discriminate between the two hypotheses
even though the underlying simulated data does not have any luminosity
evolution.

\begin{figure*}[htp] \centering \leavevmode\epsfysize=9cm
\epsfbox{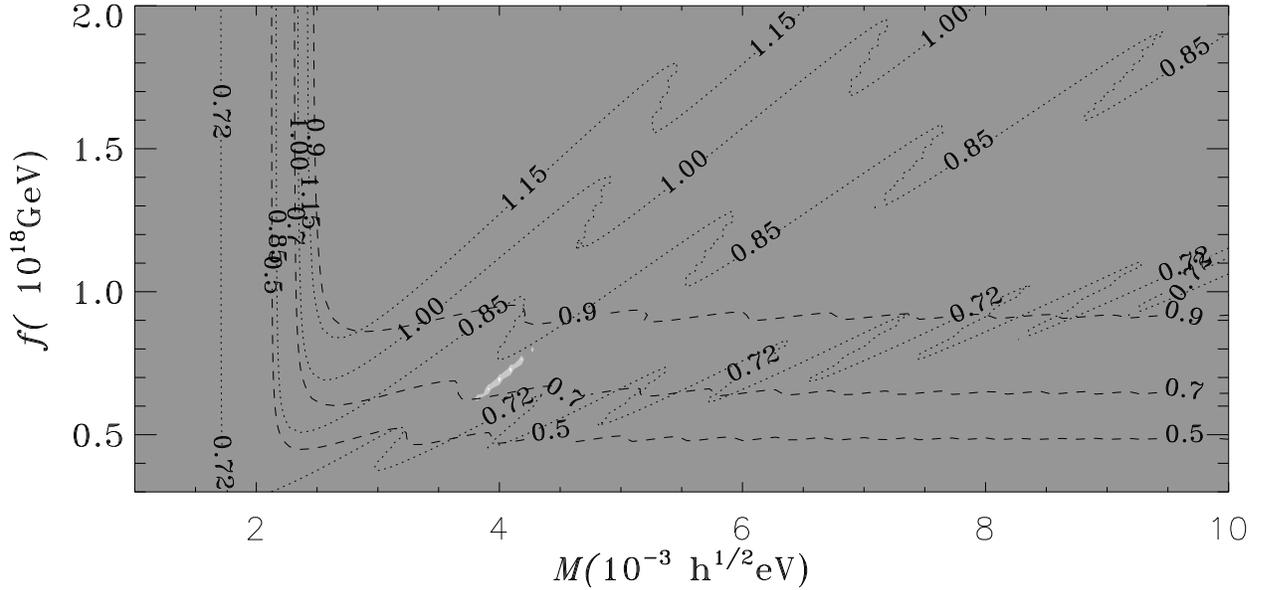} \caption[snap-evoln]{\label{snap-evoln} Confidence
limits on $M$,$f$ parameter values, with $w_i=0.2$, marginalized over a
flat prior for the luminosity evolution parameter $\bt$, for the 1915
supernovae simulated assuming $M=4\times10^{-3}h^{1/2}$eV,
$f=0.676\times10^{18}$GeV, and $\bt=0.659$.
Parameter values excluded at the 95.4\%
level are darkly shaded, while those excluded at the 68.3\% level are lightly
shaded. Overplotted are the contours for $\Omp$ (dashed) and
$H_0t_0$ (dotted).} \end{figure*}

If we fit the data set obtained from the fiducial model B by assuming that
there is no luminosity evolution, then the 95.4\% level bounds a very small
region around the best--fit values of $M\approx2.27\times10^{-3}h^{1/2}$eV and
$f\approx0.764\times10^{18}$GeV. The average parameter likelihood is ${\cal L}
\ns{ave}\simeq9.4\times10^{-18}{\cal L}_0$, where ${\cal L}_0=\exp(-1924/2)$.
If luminosity evolution is considered to exist (see Fig. \ref{snap-evoln}), on
the other hand, then the 95.4\% level bounds a small region around the
best--fit values of $M\approx4.00\times10^{-3}h^{1/2}$eV and $f\approx0.695
\times10^{18}$GeV, with $\hat\bt\approx0.626$ and normalized $\chi^2=1924$.
The average parameter likelihood is ${\cal L}\ns{ave}\simeq8.3\times10^{-4}
{\cal L}_0$. This time, the Bayes factor for the model with luminosity
evolution relative to the model without luminosity evolution is much greater
than one, $B\simeq9.1\times10^{13}$. Thus if the true data were derived from
fiducial model B, it would show up clearly as giving a $d_L(z)$ which can be
discriminated from the model without luminosity evolution.

This result in no way contradicts the inconclusive result obtained from data
set A. If the underlying data were to truly follow a simple luminosity
evolution, then the analysis above shows that this would stand out as a very
strong positive signal in the Bayes factor test. If the underlying data has no
luminosity evolution on the other hand, then it may be possible to obtain a
good fit with an extra luminosity evolution parameter, simply because one has
an extra parameter to fit. Thus a simple luminosity evolution is easy to rule
in, but difficult to rule out. Combined together the two results demonstrate
the efficacy of the Bayes factor approach. With data set B we would obtain a
decisive result, but by Ockham's razor we should reject the more complicated
evolutionary hypothesis if we were to obtain an inconclusive result
such as that pertaining to data set A. In fact, we have found that by
narrowing the range of the prior $\Delta\bt$ that the Bayes factor for
data set A can even be made slightly greater than the cut-off value of 3
listed in Table 1. Clearly any weakly positive results should therefore
also be treated with some caution.

\section{PNGB versus other potentials}

The constraints on cosmological parameters which arise from type Ia supernovae
have been broadly studied for a wide range of quintessence models
\cite{WM,Ng,NW,Wi}. However, all these models
seem to have ranges of parameters that fit the existing data equally well,
with no particular model standing out as being observationally preferred.
The lack of data and the large statistical error bars are among the
contributing reasons. In this section, we will investigate the potential of
the SNAP supernovae data set to discriminate between competing quintessence
models, as contrasted with the P98 data set, assuming no source evolution.

\begin{figure}[htp] \centering \leavevmode\epsfysize=8cm
\epsfbox{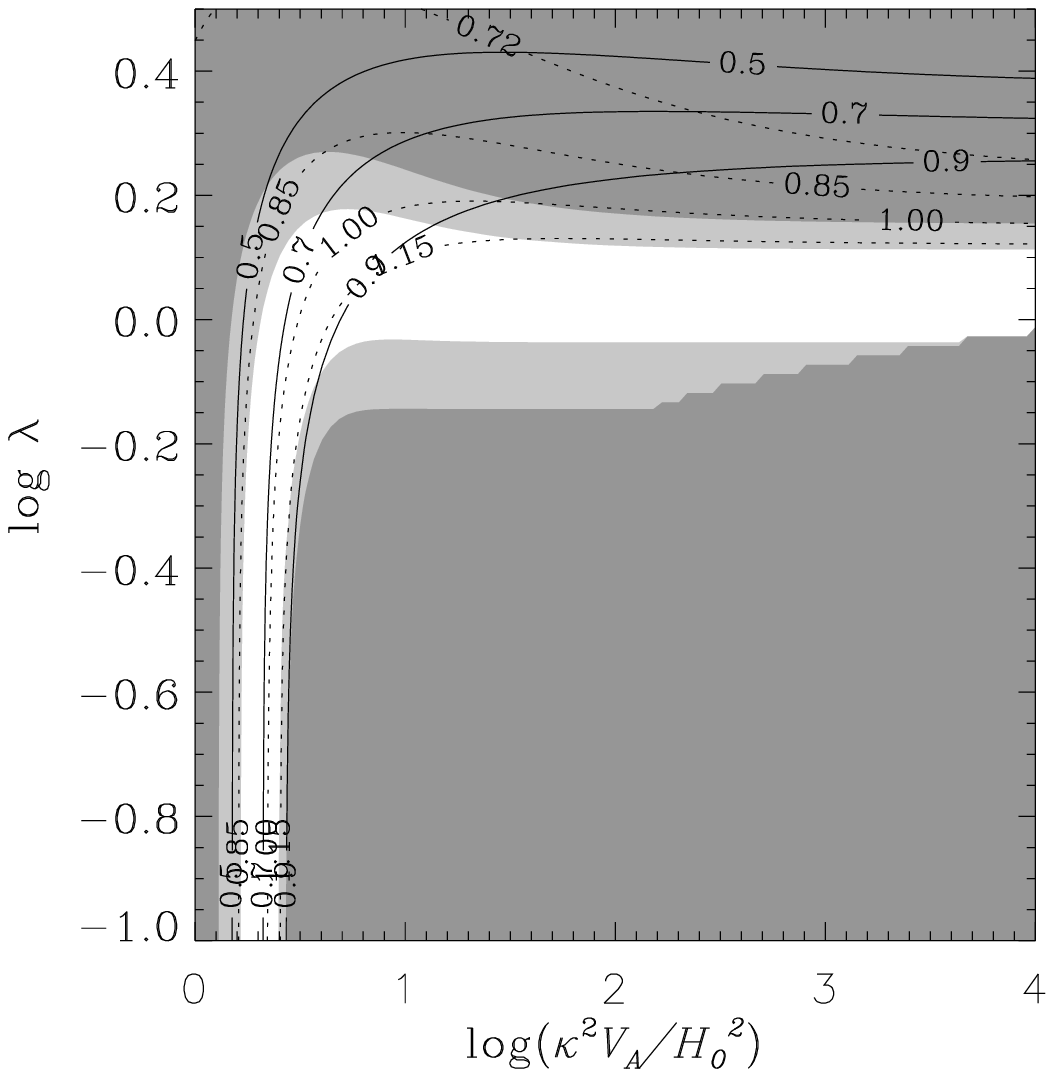} \caption[expV]{\label{expV} Confidence limits on the
parameter space for the simple exponential potential model $V=V_A\e^{-\la\kp
\ph}$, for the 54 supernovae Ia in the P98 ``fit C'' reduced data set.
Parameter values excluded at the 95.4\% level are darkly shaded, while those
excluded at the 68.3\% level are lightly shaded. Overplotted are the contours
for $\Omp$ (solid) and $H_0t_0$ (dotted).} \end{figure}

\begin{figure}[htp] \centering \leavevmode\epsfysize=8cm
\epsfbox{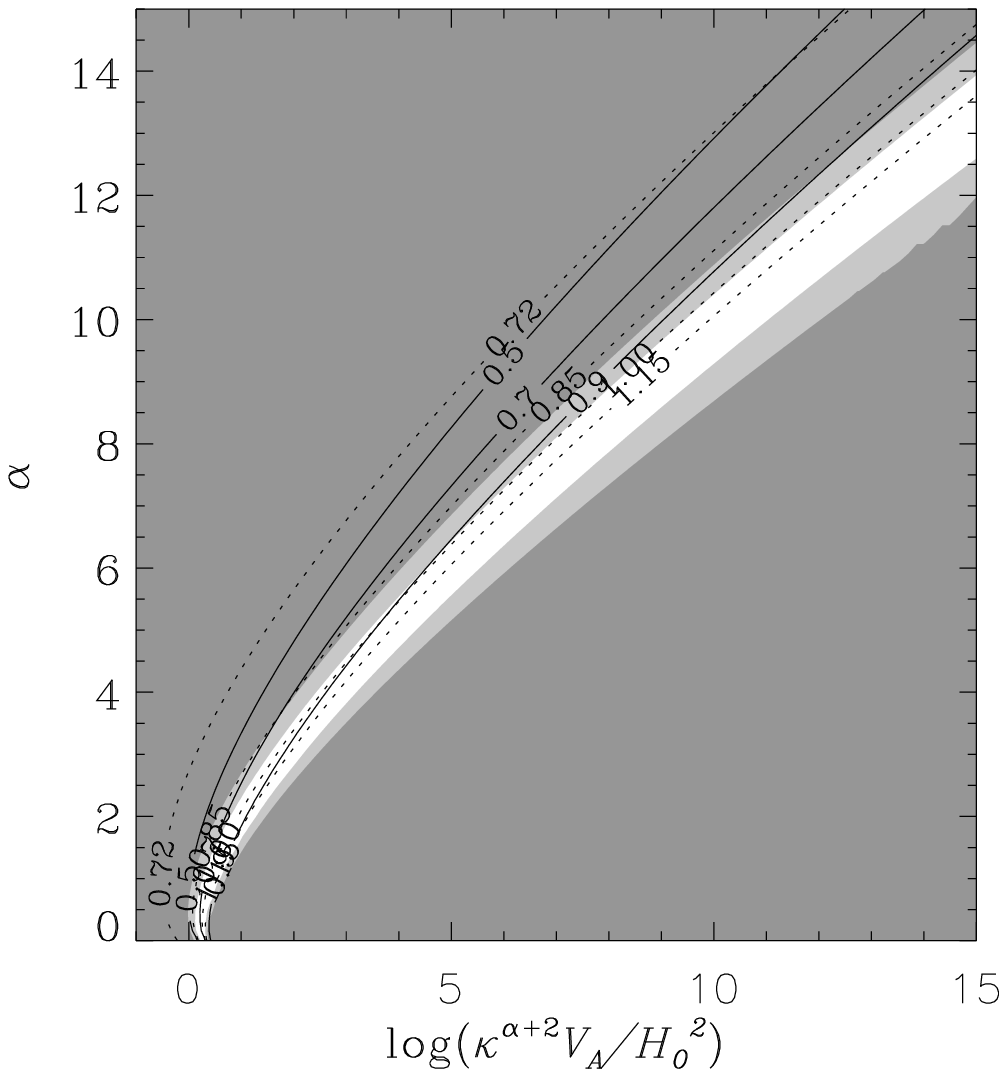} \caption[power]{\label{power} Confidence limits on the
parameter space for the inverse power--law potential model $V=V_A\ph^{-\al}$,
for the 54 supernovae Ia in the P98 ``fit C'' reduced data set. Parameter
values excluded at the 95.4\% level are darkly shaded, while those
excluded at the 68.3\% level are lightly shaded. Overplotted are the contours
for $\Omp$ (solid) and $H_0t_0$ (dotted).} \end{figure}

\begin{figure}[htp] \centering \leavevmode\epsfysize=8cm
\epsfbox{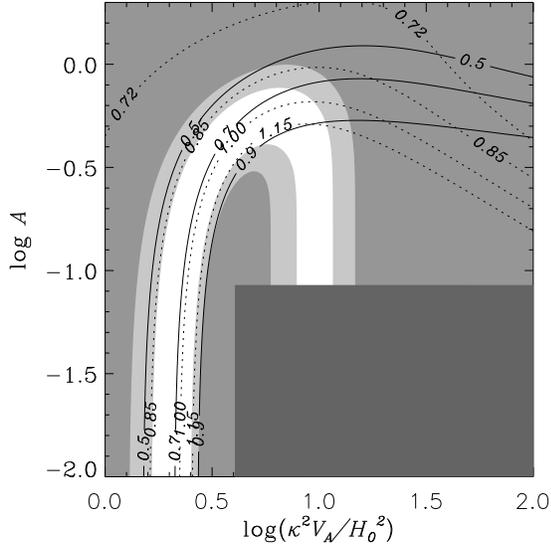} \caption[dsb]{\label{dsb} Confidence limits on the
parameter space for the double--exponential potential model $V=V_A\exp(-A\e^
{\sqrt{2}\kp\ph})$, for the 54 supernovae Ia in the P98 ``fit C'' reduced data
set. Parameter values excluded at the 95.4\% level are darkly shaded, while
those excluded at the 68.3\% level are lightly shaded. Overplotted are the
contours for $\Omp$ (solid) and $H_0t_0$ (dotted). The lower right-hand
region is excluded from the plot due to computational difficulties.}
\end{figure}

We consider some simple scalar field potential functions which have been widely
studied in the literature. They are the simple exponential potential
\cite{Wet,Ng,exp2}
\be \label{Vexp} V=V_A\e^{-\la\kp\ph}\ , \ee
the inverse power--law potential \cite{PR,power2}
\be \label{Vpower} V=V_A\ph^{-\al}\ , \ee
and the double--exponential potential \cite{Ng,dsb}
\be \label{Vdsb} V=V_A\exp(-A\,\e^{\sqrt{2}\kp\ph})\ ,\ee
where $\kp$ is related to Newton's constant by $\kp^2=8\pi G$.
We evaluate the luminosity distance -- redshift relation by numerically
evolving the Einstein-scalar field evolution equations forward from the time
of onset of matter domination, as what we did for the PNGB models. For
the simple exponential and the double--exponential potentials, without lost
of generality we choose the initial condition $\ph_i=0$. For the
power--law potential, we begin the integration by assuming that the scalar
field has already reached the tracker solution. These leave us with two
parameters in each case. In Figs. \ref{expV}, \ref{power}, and \ref{dsb},
we display the confidence limits for the 54 supernovae Ia in the P98
``fit C'' reduced data set, on the parameter spaces for these quintessence
models. (The equivalent figures for the full data set have been omitted,
as the 1$\si$ and 2$\si$ confidence regions are very similar to those shown
in Figs.\ \ref{expV}--\ref{dsb}, but just slightly narrower.)

In this section we will only compare models (\ref{Vexp})--(\ref{Vdsb})
on the basis of assuming that the true data follows the PNGB model, with
the fiducial parameters of data set A. Naturally, we could assume any of
the other potentials (\ref{Vexp})--(\ref{Vdsb}) as our fiducial model,
and then compare the other models on the basis of a new fiducial data
set. This would allow us, for example, to compare the inverse power--law
potential (\ref{Vpower}) with the simple-- and double--exponential
potentials (\ref{Vexp}), (\ref{Vdsb}), a test which we have not
performed here. However, we will restrict our attention to the fiducial
data set A relevant to the PNGB potential (\ref{VPNGB}), since the
amount of computer time involved in computing the fiducial data sets is
large, and an analysis based on a fiducial PNGB model will prove to be
sufficient to illustrate the dramatically increased discriminatory power
of a SNAP dataset as opposed to presently available SNe Ia data.
Naturally, one could extend the discussion to other fiducial data sets, but
we will leave that to future work.

\subsection{P98 data set}

Firstly we will calculate the Bayes factor of the P98 data set to compare
the PNGB model with all the other quintessence models.
In order to calculate Bayes factors which compare different theoretical
models we must naturally make choices of the range of prior parameters
integrated over, and we are dealing with different parameters in the
different models. Different choices of priors for these parameters
would lead to some variation in the resulting Bayes factors. We will make a
choice of the range of prior parameters by using the resulting values of
$\Omp$ and $H_0t_0$ that they give rise to. However, the parameter
spaces are not completely bounded by the $\Omp$ and $H_0t_0$
constraints. Without a physical cut-off for the parameters, such as a
cut-off for $f$ at Planck scale for the PNGB model, we will investigate in each
case the dependence of the Bayes factor comparison on the prior ranges of
the parameters. We will show that in some cases, the choice of the prior
parameter ranges do not affect the conclusion from the Bayes factor
analysis.

For the simple exponential potential model, as shown in Fig.\ \ref{expV},
the $\Omp$ and $H_0t_0$ contours diverge at large $V_A$, putting a bound on
the parameter space. In the small $\la$ region, where the potential is
effectively a cosmological constant model whose value depends on $V_A$ only,
the $\Omp$ and $H_0t_0$ contours remain near parallel. For the full P98 data
set the average parameter likelihood over the region bounded by the $\Omp$
and $H_0t_0$ constraints, with a cut-off at $\la=0.1$, is ${\cal L}\ns{ave}
\simeq0.67{\cal L}_R$, where ${\cal L}_R$ is the average parameter likelihood
for the PNGB model. In the small $\la$ region the likelihood only changes
slightly with decreasing $\la$ and tends to a constant, corresponding to an
average parameter likelihood which we estimate to be ${\cal L}\ns{ave}\simeq
1.5{\cal L}_R$. Since the small $\la$ region extends to an infinite range in
terms of the parameterization shown in Fig.\ \ref{expV}, if we enlarge the
prior range of $\la$ to include smaller values, then it has the effect of
bringing the average parameter likelihood closer to the average parameter
likelihood in the small $\la$ region, which is effectively an upper bound.
Thus the Bayes factor for the simple exponential potential model versus the
PNGB model lies in the range $0.67<B<1.5$ for the full P98 data set.
Similarly $0.65<B<1.5$ for the reduced ``fit C'' data set. In each case the
larger value is the one appropriate to including the entire small parameter
range for $\la$. Such a Bayes factor is too small to confidently discriminate
between the two models.

For the double--exponential potential model, the average parameter likelihood
over the region bounded by the $\Omp$ and $H_0t_0$ constraints, with a cut-off
at $A=0.01$, is ${\cal L}\ns{ave}\simeq0.92{\cal L}_R$ for the full P98 data
set. The average parameter likelihood in the small $A$ region, where the
potential is effectively a cosmological constant model whose value depends on
$V_A$ only, is ${\cal L}\ns{ave}\simeq1.5{\cal L}_R$. By a similar argument to
above, the Bayes factor for the double--exponential potential model versus the
PNGB model varies slightly over a range $0.92<B<1.5$ for the full P98 data
set. Similarly, $0.38<B<1.5$ for the reduced ``fit C'' data set. Again, the
Bayes factor is too small to confidently discriminate between the two models.

For the inverse power--law potential model, as shown in Fig.\ \ref{power},
the $\Omp$ and $H_0t_0$ contours remain nearly parallel towards the
upper right-hand region where both $V_A$ and $\al$ are large. The average
parameter likelihood over the region bounded by the $\Omp$ and
$H_0t_0$ constraints with arbitrary cut-offs at
$\kp^{\al+2}V_A/H_0^2=10^{15}$ and $\al=15$ is
${\cal L}\ns{ave}\simeq0.14{\cal L}_R$ for the full P98 data set,
and ${\cal L}\ns{ave}\simeq0.11{\cal L}_{RC}$ for the reduced ``fit C''data
set, where ${\cal L}_{RC}$ is the average parameter likelihood for the PNGB
model based on the ``fit C'' data set. This gives a Bayes factor for the
PNGB model versus the inverse power--law potential model of $B=7.2$ for the
full P98 data set, or $B=9.2$ for the ``fit C'' data set. Both values
are large enough to provide slightly positive evidence that the P98
data set favours the PNGB model over the inverse power--law potential model.
Since the Bayes factor is only weakly positive, however, we cannot not place
strong confidence in this conclusion. Note that the $\Omp$ and $H_0t_0$
contours both extend towards the smaller likelihoods region. Therefore,
increasing the prior ranges for $V_A$ and $\al$ will decrease the average
parameter likelihood, and increase the slightly positive preference for the
PNGB potential as compared to the inverse power--law potential in the
Bayes factor test.

\subsection{SNAP data set}

We have shown that the P98 data set does not particularly favour any of the
PNGB, simple exponential potential, and double--exponential potential models
relative to the others. The P98 data set does disfavour the inverse power--law
potential model as compared with the
PNGB model. This might be considered to result from the fact that the
$2\si$--confidence region only overlaps with a small region
(the low $\al$ region) of the whole parameter space allowed by the priors
set on $\Omp$ and $H_0t_0$.

In this section we want to compare the typical results we should expect when
a future SNAP data set is used. We will use data set A, simulated from a
fiducial PNGB model assuming no luminosity evolution. We will study the
potential of this data set to distinguish the PNGB model from other
quintessence models.

For the simple exponential potential model, the average parameter likelihood
over the region bounded by the $\Omp$ and $H_0t_0$ constraints, with a
cut-off at $\la=0.1$, is ${\cal L}\ns{ave}\simeq2.1\times10^{-5}{\cal L}_0$,
where ${\cal L}_0=\exp(-1910/2)$. In the small $\la$ region, we estimate the
average parameter likelihood to be ${\cal L}\ns{ave}\simeq6.6\times10^{-31}
{\cal L}_0$. Comparing with the average parameter likelihood for the PNGB
model, ${\cal L}\ns{ave}\simeq7.1\times10^{-4}{\cal L}_0$, the Bayes factor for
the PNGB model versus the simple exponential potential model takes values in
the range $34$ -- $10^{27}$, depending on the prior range for $\la$. This would
provide a very strong conclusion that data set A favours the PNGB model over
the simple exponential potential model.

For the double--exponential potential model, the average parameter likelihood
over the region bounded by the $\Omp$ and $H_0t_0$ constraints,
with a cut-off at $A=0.01$, is
${\cal L}\ns{ave}\simeq1.6\times10^{-3}{\cal L}_0$. The average parameter
likelihood in the small $A$ region, is
${\cal L}\ns{ave}\simeq2.1\times10^{-29}{\cal L}_0$. The Bayes factor for the
PNGB model versus the double--exponential potential model lies in the range
$0.44$ -- $10^{25}$, depending on the the prior range for $A$.
The lower bound of the Bayes factor, $B=0.44$, is too small to confidently
discriminate between the two models. However, provided small values of the
parameter $A$ are included in the prior range, the
Bayes factor would provide a strong conclusion that data set A favours the PNGB
model over the double--exponential potential model.

For the inverse power--law potential model, the average parameter likelihood
over the region bounded by the $\Omp$ and $H_0t_0$ constraints
with arbitrary cut-offs at $\kp^{\al+2}V_A/H_0^2=10^{15}$ and
$\al=15$ is ${\cal L}\ns{ave}\simeq1.9\times10^{-10}{\cal L}_0$. This gives
a Bayes factor for the PNGB model versus the inverse power--law potential
model of $B\simeq10^6$. This provides very strong evidence that data set A
would favour the PNGB model over the inverse power--law potential model.

\section{Discussion}

Let us now consider our results in relation to previous work concerning the
feasibility of determining the properties of the dark energy from future
supernovae surveys \cite{St,Hut,Maor,WA,BM,WG}. One approach in past studies
of scalar field quintessence has been to assume a potential $V(\ph)$ over
which the scalar field, $\ph$, would have slowly varied during cosmological
time scales, and then test the efficacy of reconstructing the potential
\cite{St,Hut}. Another approach has been to assume that the quintessence
field can be described by a perfect fluid with slowly varying equation of
state $P=w(z)\rh$, expand $w(z)$ in a power series, and then test the efficacy
of determining the coefficients in the power series \cite{Maor,WA,BM,WG}.

The conclusions of investigations to date are mixed \cite{Maor,WA,BM,WG}.
Weller and Albrecht \cite{WA} find that many models can be distinguished
with a fit to a linear
equation of state for the dark energy, $P=w(z)\rh$ with $w(z)=w_0+w_1z$, but
only if the current mass density, $\Omega_m$, is known to a high precision.
Barger and Marfatia \cite{BM} find that, even by putting
$\Omega_m=0.3$ exactly, there is still a possibility of obtaining data
sets which might not discriminate between quintessence and ``$k$-essence'',
namely an alternative form of dark energy with a scalar field characterized
by non-linear kinetic terms \cite{kessence}. Wang and Garnavich
\cite{WG} consider two classes of functions $w(z)$, corresponding
respectively to a linear variation and to $k$-essence. Using somewhat
different techniques to other authors they are more optimistic about
prospects for determining $w(z)$ from future SNe Ia data.

In the present paper we have taken a different approach from those above,
by considering a class of models -- the PNGB
models -- which are very well motivated from the point of view of particle
physics, but for which the above methods will not always be satisfactory in the
case of all plausible parameter ranges, given the potentially oscillatory
nature of $w(z)$ and the corresponding fact that the scalar field may have
varied over a wide range of values of $V(\ph)$ over observable time scales.
We fitted the simulated SNAP data sets
to the exact $d_L(z)$ of different models obtained by numerical integration,
and compared them to other models.

One real drawback of all approaches is that as yet there is no preferred
physical model for the dark energy. On one hand this means that any
approximations made in potential reconstruction methods may be too
restrictive, since many different potential energy functions
$V(\ph)$ are conceivable, and many of these may give results degenerate
with each other. On the other hand, using a given Lagrangian for the
quintessence field, as we have done, limits us to a model by model test.

Nevertheless, we find that data sets such as those that would be
produced by SNAP promise to be very successful on some tests, even if they
will probably be less successful on others. In particular, while existing
data is not yet sufficiently large to discriminate between various
quintessence models or models with evolution \cite{WM,Ng,NW},
we have shown that the much larger size and smaller error bars of the
simulated SNAP type IA supernovae data sets provide much tighter
constraints on the parameters for quintessence models such as those
corresponding to pseudo Nambu--Goldstone bosons.

By evaluating Bayes factors in the context of the PNGB model, we have shown
that future satellite SNe Ia data sets should have greater success in detecting
whether the observed luminosity distance -- redshift relation is purely
cosmological in origin, or is significantly contaminated by evolutionary
effects of the sources. The results of section IIB showed that although it may
be difficult to completely rule out luminosity evolution, if the true data were
from a population with luminosity evolution then this would provide a strong
distinctive signal.
We have only studied one simple illustrative supernova luminosity evolution
function, but we expect that similar conclusions would apply to other
simple luminosity evolution models.

We have further shown that with the future data it should be possible to
discriminate PNGB model from some other particular types of quintessence; in
particular, it gives a very different signature to that of simple inverse
power--law potentials or simple exponential potentials. The case of a
double--exponential potential gives a lower Bayes factor, and may therefore
be more difficult to distinguish from the PNGB model. However, even in this
case some distinction between the two models is possible if one allows a
suitably large prior range of the parameter $A$, to include small values.

A number of obvious extensions of our analysis are possible. In the case of
testing source evolution versus the case with no evolution, for example, it
would be interesting to determine by how much we can reduce the parameter
$\bt$ for the simulated SNAP data set B while still obtaining a very strong
result for the discriminatory power of the relevant Bayes factor. Given the
magnitude of the value obtained in section IIB, we suspect that the fiducial
value of $\bt$ could be significantly reduced. Likewise many other tests could
be performed with fiducial data sets based on other quintessence potentials.

In conclusion, we find that future supernovae measurements such as those
that would be afforded by the SNAP satellite, will have the power to
significantly increase our knowledge of the properties of the dark energy
in the universe. To be completely confident, however, we will require a deeper
theoretical understanding of the nature of the dark energy and hopefully new
input from fundamental physics.

\section*{Acknowledgements}

DLW wishes to acknowledge the financial support of Australian Research
Council grant F6960043.
\def\PRL#1{Phys.\ Rev.\ Lett.\ {\bf#1}} \def\PR#1{Phys.\ Rev.\ {\bf#1}}
\def\ApJ#1{Astrophys.\ J.\ {\bf#1}} \def\AsJ#1{Astron.\ J.\ {\bf#1}}
\def\CQG#1{Class.\ Quantum Grav.\ {\bf#1}} \def\Nat#1{Nature {\bf#1}}
\def\JMP#1{J.\ Math.\ Phys.\ {\bf#1}} \def\NP#1{Nucl.\ Phys.\ {\bf#1}}
\def\PL#1{Phys.\ Lett.\ {\bf#1}} \def\AsAp#1{Astron.\ Astrophys.\ {\bf#1}}
\def\MNRAA#1{Mon.\ Not.\ Roy.\ Astron.\ Soc.\ {\bf#1}}

\end{document}